\documentclass[prl,floatfix,twocolumn,showpacs]{revtex4}
\usepackage{amsmath}
\usepackage{amsfonts}
\usepackage{mathrsfs}
\usepackage{epsfig}

\usepackage{graphicx}
\usepackage{dcolumn}
\usepackage{amsmath}
\usepackage{color}

\def\rev2#1{{\color{red}{ #1}}}

\def\be{\begin{equation}}
\def\ee{\end{equation}}

\begin{document}
\title{
Self-organized transition to coherent activity in disordered
media
}
\author{Rajeev Singh$^1$, Jinshan Xu$^{2,3}$, Nicolas Garnier$^2$, Alain
Pumir$^2$ and Sitabhra Sinha$^1$}
\affiliation{
$^1$The Institute of Mathematical Sciences, CIT Campus, Taramani,
Chennai 600113, India.\\
$^2$ Laboratoire de Physique, ENS de Lyon and CNRS, 46
All\'ee d'Italie, 69007, Lyon, France. \\
$^3$Department of Physics, East China
Normal University, Shanghai, 20062, China.
}
\date{\today}
\begin{abstract}
Synchronized oscillations are of critical functional importance in
many biological systems. We show that such oscillations can arise
without centralized coordination in a disordered system of
electrically coupled excitable and passive cells. Increasing the
coupling strength results in waves that lead to coherent periodic
activity, exhibiting cluster, local and global synchronization under
different conditions.  Our results may explain the self-organized
transition in a pregnant uterus from transient, localized activity
initially to system-wide coherent excitations just before delivery. 
\end{abstract}

\maketitle

\newpage
Rhythmic behavior is central to the normal functioning of many
biological processes~\cite{Glass01} and 
the periods of such oscillators span a wide range of time scales
controlling almost every aspect of
life~\cite{Gillette05,Golubitsky99,Winfree00,Hakim09}.
Synchronization of spatially distributed oscillators is of crucial
importance for many biological systems~\cite{Pikovsky03}. 
For example, disruption of
coherent collective activity in the heart can result
in life-threatening arrhythmia~\cite{Keener98}.
In several cases, the rhythmic behavior of the entire system is
centrally organized by a specialized group of oscillators (often
referred to as {\em pacemakers})~\cite{Chigwada06} as in the heart,
where this function is performed in the sino-atrial node~\cite{Tsien79}.
However, no such special coordinating agency has been identified for
many biological processes.
A promising mechanism
for the self-organized emergence of coherence is
through coupling among neighboring elements. 
Indeed, local interactions can lead to order without an
organizing center in a broad class of complex
systems~\cite{Gregoire04}.

The present work is inspired by studies of the pregnant uterus whose
principal function is critically dependent on
coherent rhythmic contractions that, unlike the heart,
do not appear to be centrally coordinated
from a localized group of pacemaker cells~\cite{Blackburn07}. In fact,
the uterus remains quiescent almost throughout pregnancy
until at the very late stage when large sustained periodic activity is
observed immediately preceding the expulsion of the
fetus~\cite{Garfield07}.
In the USA,
in more than $10~\%$ of all pregnancies,
rhythmic contractions are initiated significantly earlier, causing
preterm births~\cite{Martin09}, which are responsible 
for more than a third of all 
infant deaths~\cite{MacDorman07}.
The causes of premature rhythmic activity are not well
understood and at present there is no effective
treatment for preterm labor~\cite{Garfield07}.

\begin{figure}
 \centering
 \includegraphics[width=0.99\linewidth,clip]{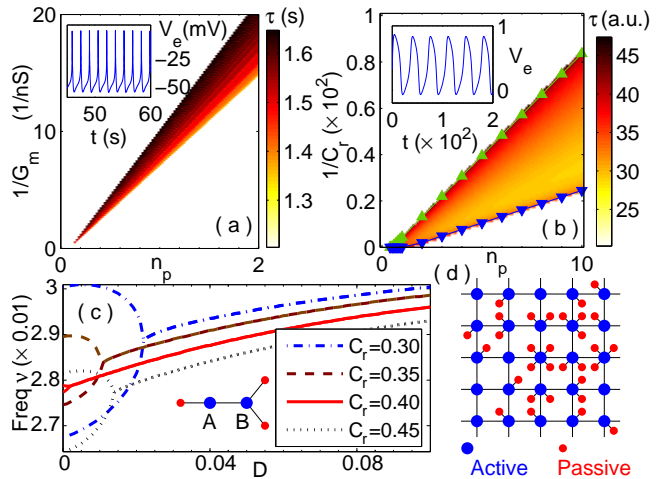}
 \caption{(Color online) 
Oscillations through interaction between excitable and
passive elements. A single excitable element described by (a) a
detailed ionic model of an uterine myocyte and (b) a generic FHN
model, coupled to $n_p$
passive elements
exhibits oscillatory activity (inset) with period $\tau$
for a specific range of gap junctional conductances $G_m$ in (a)
and coupling strengths $C_r$ in (b). The triangles (upright and
inverted) enclosing 
the region
of periodic activity in (b) are obtained analytically 
by linear stability
analysis of the fixed
point solution of Eq.~(\ref{eqn3a}).
(c) Frequency of oscillation for a system of two ``oscillators" $A$ and
$B$ (each comprising an excitable cell and $n_p$ passive cells with
$n_p^A = 1$ and $n_p^B = 2$) coupled 
with strength $D$. Curves corresponding to
different values of $C_r$ show that the system synchronizes on
increasing $D$, having a frequency that can be {\em higher} than either 
of the component oscillators.
(d) Uterine tissue model as a 2-dimensional square lattice,
every site occupied by an excitable cell coupled
to a variable number of passive cells.
 }
\label{fig1}
\end{figure}

In this paper we have investigated the emergence of coherence using a
modeling approach that stresses the role
of coupling in a system of heterogeneous entities.
Importantly, recent studies have not revealed the presence of 
pacemaker cells in the uterus~\cite{Shmygol07}.
The uterine tissue has a heterogeneous composition, comprising
electrically excitable smooth muscle cells (uterine myocytes),
as well as electrically passive cells (fibroblasts and interstitial
Cajal-like cells [ICLCs])~\cite{Duquette05,Popescu07}.
Cells are coupled in tissue by gap junctions that serve as
electrical conductors. In the uterine tissue,
the gap junctional couplings have been seen to markedly increase
during late pregnancy and labor, both in 
terms of the number of such junctions and their
conductances (by an order of magnitude~\cite{Miller89}), 
which is the most striking
of all electrophysiological changes the cells undergo during this period.
The observation that isolated uterine cells do not spontaneously 
oscillate~\cite{Shmygol07}, 
whereas the organ rhythmically contracts when the number of gap junctions 
increases, strongly suggests a prominent role of the 
coupling.
The above
observations have motivated our model for the onset
of spontaneous oscillatory activity and its synchronization through
increased coupling in a
mixed 
population of excitable and passive elements.
While it has been shown earlier that an excitable cell
connected to passive cells can oscillate~\cite{Jacquemet06}, 
we demonstrate that
coupling such oscillators with different
frequencies (because of varying numbers of passive cells) can result in
the system having a frequency {\em higher} than its
constituent elements. 
We have also performed a systematic characterization for the first time of
the dynamical transitions occurring in the heterogeneous medium
comprising active and passive cells
as the coupling is increased, revealing a rich variety of
synchronized activity in the absence of any pacemaker.
Finally, we show that the system 
has multiple coexisting attractors characterized by 
distinct mean oscillation periods, with the nature of variation of
the frequency with coupling depending on the choice of
initial state as the coupling strength is varied.
Our results provide a
physical understanding of the transition from transient excitations to
sustained rhythmic activity through physiological changes such as
increased gap junction expression~\cite{Garfield98}. 
The dynamics of excitable myocytes can be described by a model having
the form $C_m \dot{V_e} = -I_{ion}(V_e, g_i)$ where $V_e$(mV) is the
potential difference across a cellular membrane, $C_m$~(= 1 $\mu$F
cm$^{-2}$) is the membrane capacitance, $I_{ion}$~($\mu$A~cm$^{-2}$)
is the total current density through ion channels on the cellular
membrane and $g_i$ are the gating variables, describing the 
different ion channels.
The specific functional form for $I_{ion}$ varies in different models.
To investigate the actual biological system we have first considered a
detailed, realistic description of the uterine myocyte given by Tong
{\em et al.}~\cite{Tong11}. However, during the systematic dynamical 
characterization of the spatially extended system, for ease of
computation we have used the phenomenological FitzHugh-Nagumo (FHN)
system~\cite{Keener98} which exhibits behavior qualitatively
similar to the uterine myocyte model in the excitable regime.
In the FHN model, the ionic current is given by
$I_{ion} = F_e (V_e,g) = A V_e (V_e-\alpha) (1-V_e) - g$, 
where $g$ is an effective
membrane conductance evolving with time as $\dot{g} = \epsilon
(V_e - g)$, $\alpha (=0.2)$ is the excitation threshold, $A
(=3)$ specifies the fast activation kinetics and $\epsilon (=0.08)$
characterizes the recovery rate of the medium (the parameter values
are chosen such that the system is in the excitable regime
and small variations do not affect the results
qualitatively).
The state of the electrically passive cell is 
described by the time-evolution of the single variable $V_p$~\cite{Kohl94}:
$\dot{V_p} = F_p(V_p)=K(V^R_p-V_p)$,
where the resting state for the cell, $V^R_p$ is set to 1.5 and $K
(=0.25)$ characterizes the time-scale over which perturbations away from
$V^R_p$ decay back to it.
We model the interaction between a myocyte and one or more
passive cells by:
\begin{subequations}
\begin{align}
\dot{V_e} &= F_e(V_e,g)+n_p\ C_r (V_p-V_e), 
\label{eqn3a} \\
\dot{V_p} &= F_p(V_p)-C_r (V_p-V_e),
\label{eqn3}
\end{align}
\end{subequations}
where $n_p (= 1, 2, \ldots)$ passive elements are
coupled to an excitable element via the
activation variable $V_{e,p}$ with
strength $C_r$. Here, we have assumed for simplicity that all passive
cells are identical having the same parameters $V^R_p$ and $K$, as
well as, starting from the same initial state.
We observe that the coupled system comprising a realistic model of uterine
myocyte and one or more passive cells exhibits oscillations
(Fig.~\ref{fig1}~(a)) qualitatively similar to the generic FHN model
(Fig.~\ref{fig1}~(b)), although the individual elements
are incapable of spontaneous periodic activity in both cases.
In Fig.~\ref{fig1}~(a-b), the range of $n_p$ and excitable-passive
cell couplings for which limit cycle oscillations of the coupled system
are observed is indicated with a pseudocolor representation of the
period ($\tau$).
We also look at how a system obtained by diffusively coupling two such
``oscillators" with distinct frequencies (by virtue of having
different $n_p$) behaves upon increasing the coupling constant $D$
between $V_e$
(Fig.~\ref{fig1}~(c)). A surprising result here is that the combined
system may oscillate {\em faster} than the individual oscillators
comprising it.  

To investigate the onset of spatial organization of periodic activity
in the system we have considered a 2-dimensional medium of locally
coupled excitable cells, where each excitable cell is connected to
$n_p$ passive cells [Fig.~\ref{fig1}~(d)], $n_p$ having a Poisson
distribution with mean $f$. Thus, $f$ is a measure of the density of
passive cells relative to the myocytes. Our results reported here are
for $f=0.7$; we have verified for various values of $f~\geq~0.5$ 
that qualitatively
similar behavior is seen. The dynamics of the resulting
medium is described by:
\[ \frac{\partial V_e}{\partial t} = F_e(V_e,g) + n_p\ C_r (V_p-V_e) + 
D \nabla^2 V_e, \]
where $D$ represents
the strength of coupling between excitable elements (passive cells are
not coupled to each other).
Note that, in the limit of large $D$ the behavior of the 
spatially extended medium can be reduced by a mean-field
approximation to a single excitable element coupled to $f$ passive
cells. As $f$ can be non-integer, $n_p$ in the mean-field limit 
can take fractional values [as in Fig.~\ref{fig1}~(a-b)].

We discretize the system on a square spatial grid of size $L \times L$
with the lattice spacing set to $1$. For
most results reported here $L=64$, although we have used $L$ up to
$1024$
to verify that the qualitative nature of the
transition to global synchronization with increasing coupling is
independent of system size.
The dynamical equations are solved using a fourth-order Runge Kutta
scheme with time-step $dt \leq 0.1$ and a standard 5-point stencil 
for the
spatial coupling between the excitable elements.
We have used periodic boundary conditions in the
results reported here and verified that no-flux boundary
conditions do not produce qualitatively different 
phenomena. Frequencies of individual elements are calculated using FFT
of time-series for a duration $2^{15}$ time units.
The behavior of the model for a specific set of values of $f$, $C_r$ and $D$
is analyzed over many ($\sim 100$) 
realizations of the $n_p$ distribution with random
initial conditions.

\begin{figure}
\centering
\includegraphics[width=0.99\linewidth, clip]{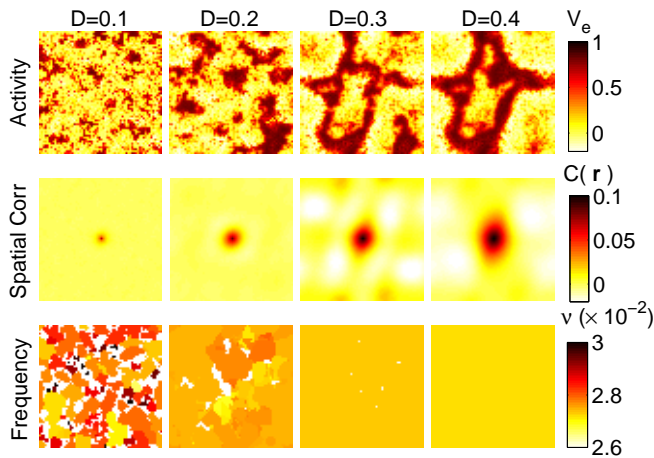}
\caption{(Color online)
Emergence of synchronization with increased coupling.
Snapshots (first row) of the activity $V_e$ in a two-dimensional 
simulation domain ($f=0.7, C_r=1, L=64$) for increasing values of 
coupling $D$ (with a given distribution of
$n_p$).
The corresponding time-averaged spatial correlation functions C({\bf
r}) are shown in the middle row. 
The size of the region around {\bf r} $= 0$ (at center) where C({\bf
r}) is high
provides a measure of the correlation
length scale which is seen to increase with $D$.
The last row shows pseudocolor plots indicating the frequencies of
individual oscillators in the medium 
(white corresponding to absence of oscillation).
Increasing $D$
results in decreasing the number of clusters with 
distinct oscillation frequencies,
eventually leading to global synchronization characterized by
spatially 
coherent, wavelike excitation patterns where
all elements in the domain oscillate with same frequency.
}
\label{fig2}
\end{figure}
Fig.~\ref{fig2}~(first row) shows spatial activity in the system
at different values of $D$ after long durations ($\sim 2^{15}$ time
units)
starting from random initial conditions.
%
As the coupling $D$ between the excitable
elements is increased, we observe a transition from highly
localized, asynchronous excitations to spatially organized coherent
activity that manifests as propagating waves. Similar traveling waves
of excitation have indeed been experimentally observed {\em in vitro}
in myometrial tissue from the pregnant uterus~\cite{Lammers08}.
The different dynamical regimes observed during the
transition are accompanied by an increase in spatial
correlation length scale~(Fig.~\ref{fig2}, middle row) and can be
characterized by the spatial variation
of frequencies of the constituent elements
(Fig.~\ref{fig2}, last row).
For low coupling ($D=0.1$), multiple clusters each with a distinct
oscillation frequency $\nu$ coexist in the medium.
As all elements belonging to one cluster have the same period, 
we refer to this behavior as {\em cluster synchronization} (CS). 
Note that there are also
quiescent regions of non-oscillating elements indicated in
white.
With increased coupling the clusters merge, reducing the
variance of the distribution of oscillation frequencies
eventually resulting in a single frequency for all
oscillating elements (seen for $D=0.3$). As there are
still a few local regions of inactivity, we term this behavior as {\em
local synchronization} (LS). 
Further increasing $D$ (=0.4), a single wave traverses the entire
system resulting in {\em global synchronization} (GS) characterized by
{\em all} elements in the
medium oscillating at the same frequency.
Our results thus help in causally connecting two well-known
observations about electrical activity in the pregnant uterus: (a)
there is a remarkable increase in cellular coupling through gap
junctions close to onset of labor~\cite{Miller89} and (b) excitations
are initially infrequent and irregular, but gradually become
sustained and coherent towards the end of labor~\cite{Blackburn07}.

\begin{figure}
 \centering
 \includegraphics[width=0.99\linewidth,clip]{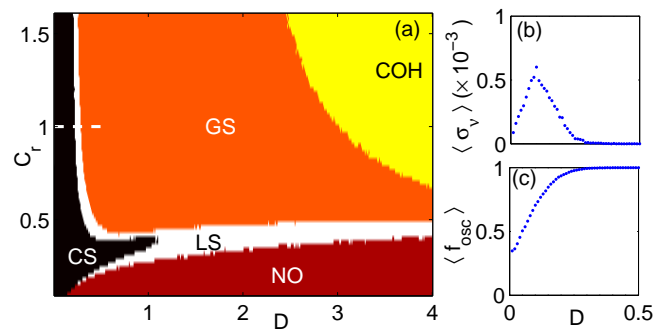}
 \caption{
 (Color online)
(a) Different dynamical regimes of the uterine tissue 
model (for $f=0.7$) in $D-C_r$ parameter plane indicating
the regions having (i) complete absence of oscillation (NO),
(ii) cluster synchronization (CS), (iii) local synchronization (LS),
(iv) global synchronization (GS) and (v) coherence (COH).
(b-c) Variation of (b) width of frequency distribution $\langle
\sigma_{\nu} \rangle$ and (c) fraction of oscillating cells
$\langle f_{osc} \rangle$ with coupling strength $D$ for
$C_r = 1$ [i.e., along the broken line shown in (a)]. 
The regimes in (a) are distinguished by thresholds applied on order
parameters $\langle \sigma_{\nu} \rangle$, $\langle f_{osc} \rangle$
and $\langle F \rangle$, viz., NO: $\langle
f_{osc} \rangle< 10^{-3}$, CS:
$\langle \sigma_{\nu} \rangle> 10^{-4}$, LS:
$\langle \sigma_{\nu} \rangle< 10^{-4}$ and $\langle
f_{osc} \rangle< 0.99$; GS:
$\langle f_{osc} \rangle> 0.99$ and COH: $\langle
F \rangle> 0.995$.
Results shown are averaged over many realizations.
} 
\label{fig3}
\end{figure}
The above observations motivate the following order parameters that
allow us to quantitatively segregate the different synchronization
regimes in the space of model parameters [Fig.~\ref{fig3}~(a)]. 
The CS state is
characterized by a finite width of the frequency distribution 
as measured by the standard deviation, $\sigma_{\nu}$, and the
fraction of oscillating elements in the medium, $0<f_{osc}<1$.
Both LS and GS states have $\sigma_{\nu} \rightarrow 0$, but differ in
terms of $f_{osc}$ ($<1$ in LS, $\simeq 1$ in GS). Fig.~\ref{fig3}
(b-c) shows the variation of
the two order parameters $\langle \sigma_{\nu}
\rangle$ and $\langle f_{osc} \rangle$ with the coupling $D$,
$\langle~\rangle$ indicating ensemble average over many realizations.
Varying the excitable cell-passive cell coupling $C_r$ together with
$D$ allows us to explore the rich variety of spatio-temporal behavior
that the system is capable of [Fig.~\ref{fig3}~(a)]. In addition to
the different synchronized states (CS, LS and GS), we also observe a
region where there is no oscillation (NO) characterized by $f_{osc}
\rightarrow 0$, and a state where all elements oscillate with the same
frequency and phase which we term coherence (COH). COH is
identified by the condition that the order parameter
$F \equiv {\rm max}_t [f_{act} (t)] \rightarrow 1$ where $f_{act} (t)$
is the fraction of elements that are active ($V_e > \alpha$) at time $t$.
%
%
In practice, the different states are characterized by thresholds
whose specific values do not affect the qualitative nature of the
results.

\begin{figure}
 \centering
 \includegraphics[width=0.99\linewidth,clip]{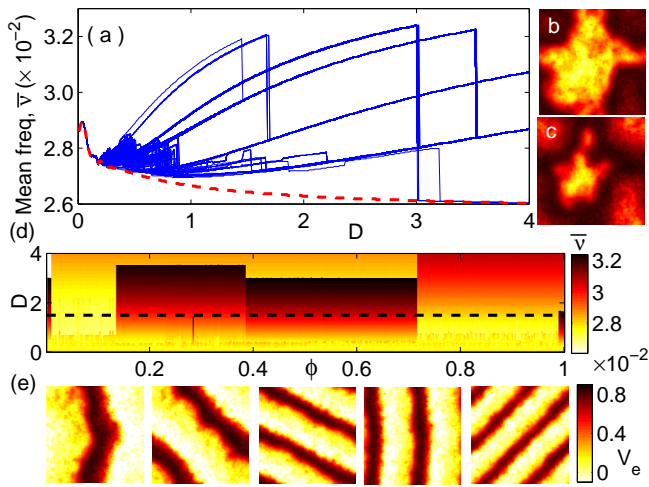}
 \caption{(Color online) 
(a) Variation of mean oscillation frequency $\bar{\nu}$ with coupling
strength $D$ in the uterine tissue model ($f = 0.7$) \
for 400 different initial conditions at $C_r = 1$.
Continuous curves correspond to gradually increasing $D$ starting from a
random initial state at low $D$, while broken curves
(overlapping)
correspond to random initial conditions chosen at each value of $D$.
(b-c) Snapshots of activity in the medium at $D=1.5$ for a
random initial condition
seen at intervals of $\delta T =5$  time units.
(d) Variation of the cumulative fractional volumes $\phi$ of the basins 
for different attractors
corresponding to activation patterns shown in (b-c) and (e), as a
function of the coupling strength $D$. (e) Snapshots of topologically
distinct patterns of activity
corresponding to the five attractors at $D = 1.5$ (broken line in (d)) 
when $D$ is increased.
}
\label{fig4}
\end{figure}
To further characterize the state of the system,
we determined the mean frequency $\bar{\nu}$ by averaging over all
oscillating cells for any given realization
of the system.
Fig.~\ref{fig4}~(a) reveals that several values of the mean frequency
are possible at a given coupling strength.
When the initial conditions are chosen randomly for each value of the
coupling (broken curve in Fig.~\ref{fig4}~(a)), the mean frequency
decreases with increasing $D$. On the other hand,
$\bar{\nu}$ is observed to {\em increase} with $D$ when
the system is allowed to evolve starting from a random initial state
at low $D$, and then adiabatically increasing the value of $D$.
The abrupt jumps correspond to drastic changes in the size of the
basin of
an attractor at certain values of the coupling strength, which can be
investigated in detail in future studies.
This suggests a multistable attractor landscape of the system
dynamics, with the basins of the multiple attractors shown in
Fig.~\ref{fig4}~(d) [each
corresponding to a characteristic spatiotemporal pattern of activity
shown in Fig.~\ref{fig4}~(e)] having differing sizes. 
They represent one or more plane waves propagating in the medium 
and are quite distinct
from the disordered patterns of spreading activity
(Fig.~\ref{fig4}~(b-c)) seen when random initial conditions are used at
each value of $D$. 
We note that
the period of recurrent activity in the uterus
decreases with time as it comes closer to term~\cite{Garfield98}
in conjunction with the increase in number of gap junctions. 
This is consistent with our result in Fig.~\ref{fig4}(a) when
considering a gradual increase of the coupling $D$.

Our results explain several important features
known about the emergence of contractions in uterine tissue.
Previous experimental results have demonstrated that the
coupling between cells in the myometrium increases
with progress of pregnancy~\cite{Miller89}.
This suggests that the changes in the system with time amounts to
simultaneous increase of $D$ and $C_r$, eventually leading to
synchronization as shown in Fig.~\ref{fig3}~(a).
Such a scenario is supported by experimental evidence that
disruption of gap-junctional communication is associated with 
acute inhibition of spontaneous uterine contractions~\cite{Tsai98}.
The mechanism of synchronization discussed here is based on a very 
generic model, suggesting that our results apply to a broad class of
systems comprising coupled excitable and passive
cells~\cite{Bub02,Pum05}.
A possible extension will be to investigate the
effect of long-range connections~\cite{Falcke94}.

To conclude, we have shown that coherent periodic activity can emerge in a
system of heterogeneous cells in a  self-organized manner and does not
require the presence of a centralized coordinating group of
pacemaker cells. A rich variety of collective behavior is observed in the
system under different conditions; in
particular, for intermediate cellular coupling, groups of cells spontaneously
form clusters that oscillate at different frequencies. With increased
coupling, clusters merge and eventually give rise to a globally
synchronized state marked by the genesis of propagating waves of excitation in
the medium. Our model predicts that a similar set of changes occur in the
uterus during late stages of pregnancy.

This research was supported in part by IFCPAR (Project 3404-4). We
thank HPC facility at IMSc for providing computational resources.

\end{document}